# A Numerical Model of Philippine Population Growth: Child Policy, Quantitative Insights and Challenges

**Dylan Antonio SJ. Talabis**
ORCID No. 0000-0002-4468-688X
dylantalabis@gmail.com
University of the Philippines Los Baños,
Los Baños, Laguna, Philippines

**Erick Justine V. Manay**
ORCID No. 0000-0001-5898-7539
ejvmanay@gmail.com
University of the Philippines Los Baños
Los Baños, Laguna, Philippines

**Ariel L. Babierra**
ORCID No. 0000-0002-6724-913X
albabierra@upd.edu.ph
University of the Philippines Los Baños
Los Baños, Laguna, Philippines

**Jabez Joshua M. Flores**
ORCID No. 0000-0002-4077-1104
jabezjoshua.flores@upou.edu.ph
University of the Philippines Open University
Los Baños, Laguna, Philippines

**Jomar F. Rabajante**
ORCID No. 0000-0002-0655-0893
jfrabajante@upd.edu.ph
University of the Philippines Los Baños
Los Baños, Laguna, Philippines

*Abstract*—The study investigates the effect of imposing an n-child policy by forecasting the population of the Philippines using a discrete age-structured compartmental model. Based on the results of the projection, a policy promoting a maximum of two children per couple leads to a transient stabilization (i.e., the population eventually declines after attaining zero-growth rate). A three-child policy may also lead to stabilization yet may converge beyond the calculated Verhulstian carrying capacity of approximately 200M. However, overshooting the carrying capacity can be resolved by increasing the available resources that can support the escalating population size. A child policy dictating a maximum of four or more children per couple results to a similar population growth as the status quo due to the inherent declining birth rate. With the declining birth rate trend in the Philippines, population stabilization is realizable even without implementing a child policy but only after 100 years. Furthermore, this study estimated the future age structure and the resultant GDP per capita income associated with each child policy.

## INTRODUCTION

Population growth and population structures are key factors for sustainable development. Several countries have already expressed their concerns about high population growth and have already started formulating reproductive health policies (UNFPA, 2012). Disproportionate population growth puts pressures on natural resources, human well-being, and global warming (Bremner et al., 2010). Whereas, developed countries and some middle income economies are raising concerns regarding declining population growth rates. This scenario led to shrink in working-age populations, rapid population ageing, and problems in renewability of the labor force and sustainability of social security and health care systems (UN DESA, 2011). As such, government policies and programs should address population stabilization and management while maintaining a balance between population processes (fertility, mortality and migration) and socio-economic development outcomes.

In 2011, the Philippines ranked as the 12th country with the highest population (CIA, 2013). With the exception of Nigeria, the population growth rate of the Philippines exceeds the rates of the 11 most populous countries (UN DESA, 2012). The declining growth rate of the Philippines is still higher than the average global population growth rate of 1.19% (UN DESA, 2012; POPCOM, 2013). The Philippines will even have a faster population growth than India and China (Jones, 2013). In Southeast Asia, the Philippines is only second to Indonesia in terms of population size. However, in the next 20 years, the greatest relative increase in population in Southeast Asia will be in the Philippines. The continued positive population growth rate and its slow decline in the Philippines are due to the continued relatively high total fertility rate (Pastrana and Harris, 2011; NSO, 2012a). High fertility is due to unwanted childbearing, and a desired family size of more than two children (Bongaarts and Bruce, 1998).

As one of the countries that signed the 1967 UN Declaration of Population and the 1994 Statement of Population Stabilization of World Leaders, the Philippines recognized the population problem "as a principal element in long-range planning if governments are to achieve their economic goals and fulfill the aspirations of their people" (Concepcion, 1973). Since 1969, the Philippine government under different administrations adopted population policies and programs of varying degrees to achieve population stabilization. However, most of these policies and programs address reproductive health and family planning that indirectly affect population growth (Herrin, 2002).

Population policies and programs in the Philippines are still in the context of the 1987 Philippine Constitution which explicitly guarantees "the right of couples to form their family and decide freely on the number of their children based on their religious beliefs and the demands of responsible parenthood" (Osias et al., 2010). A review by the Philippines Institute of Development Studies explained that the Philippine government under different administrations did not have continuity in population policies that address fertility and population growth (Herrin, 2002).

DA Talabis, EJV Manay, AL Babierra, JJM Flores, JF Rabajante (2013)

In the 1970's, the Philippine government under President Marcos launched the National Population Program to reduce the country's fertility rate. The program institutionalized the Commission on Population (POPCOM) to study population problems. Under the Aquino administration in the 1990's, population programs focused on family planning as a family health program instead of directly reducing fertility (Senate, 2009). This trend continued in the Ramos administration.

In an attempt to reduce fertility, the Estrada administration introduced alternative demographic scenarios and different contraceptive methods. However, in the next administration under Arroyo, population policies mostly reflected the position of the Catholic Church (Herrin, 2002). Government programs focused on responsible parenting, informed choice in family size, respect for life, and birth spacing. The national government advocated natural family planning as the acceptable mode of birth control. The local government oversaw other family planning methods, such as the use of contraceptives.

Population policies and programs are contentious issues in the Philippines. The congress enacted the Responsible Parenthood and Reproductive Health Act of 2012 (Republic Act No. 10354), more popularly known as the RH Bill, in December 2012. However, due to petitions challenging the law's constitutionality and criticisms from religious groups, the Supreme Court delayed its implementation in March 2013 (Lopez and Alvarez, 2013). As of August 2013, the implementation of the law is still pending (Avendaño, 2013).

The RH Bill guarantees the use of universal access to reproductive health services and supplies including family planning, contraceptives, health and sexuality education, prenatal care, and maternal care (Senate, 2009). Although the focus of RH Bill is to address total family health, it has implications to population growth and fertility. The early version of RH Bill (Section 20) declares that the State shall assist in achieving the desired size of families but will encourage families to have two children. However, this is neither mandatory nor compulsory, and there is no punitive action for having more than two children (Congress, 2013). On the average, having two children is ideal for replacing the ageing parents in the future. For most cases in the Philippines, researchers believe that there is an association between poverty and household size (Virola and Martinez, 2007).

**Overview of child policy**

In the 1960's, as a solution to the post-World War II baby boom, the Singaporean government encouraged families to have at most two children. The popular name of this program is "Stop at Two" or "Two is Enough" policy. The state offered practical incentives for having at most two children, and penalized married couples with more than two children. The program was able to reduce the fertility rate of Singapore from five births per woman in the 1960's to a replacement level in 1975 (Wong and Yeoh, 2003; Yap, 2003). The population growth rate of Singapore decreased from 4.69% in the 1950's to 1.52% in the 1970's (UN DESA, 2012). In the 1980's, the fertility rate of Singapore fell below replacement level. As a solution to this, Singapore launched the program dubbed as "Have Three or More Children If You Can Afford It" in 1987. The state encouraged unmarried singles to build a family, and initiated campaigns about the importance of family life (Wong and Yeoh, 2003). In 2001, the Singaporean government implemented the Children Development Co-Savings Scheme (or the Baby Bonus Scheme). The scheme gives procreation incentives to "lighten the financial burden of raising children" (Yap, 2003).

The People's Republic of China had a population of 580 million in the 1960's with a growth rate of 2.7% and a fertility rate of 6 births per woman (UN DESA, 2012). To reduce its growing population, the government emphasized the principle embodied in *"wan, xi, shao"* which means "later" (for delayed marriages), "longer" (for birth spacing) and "fewer" (for the number of births). The state also campaigned for "One is not too few; two are ideal; three are too many" principle (Stolc, 2008). With a goal of reducing its population, China introduced the well-known "one-child policy" in 1978 by giving incentives to one-child families, and disincentives for families having two or more children (Stolc, 2008). China was able to reduce its annual growth rate to 0.68% (1996-2000) from 2.2% (1970-1975), and the fertility rate to 2.05 (1996-2000) from 4.76 (1970-1975) (UN DESA, 2012). In 1996, the fertility rate of China fell below replacement level. The "one-child policy" resulted to an ageing population, labor shortages, gender disparities, and the so-called "4-2-1 Problem", where in one adult offspring would be left to provide support for two parents and four grandparents (Mian, 2007). Drastic declines in population growths are not distinct to Singapore and China. Other developed countries and middle income countries are also experiencing below-replacement fertility levels and declining population size.

## FRAMEWORK

Escalating population size can be detrimental when resources are scarce. Nevertheless, population growth can essentially lead to positive impacts in economic development (Simon, 1981). Followers of the "Neutralist Theory" suggest focusing on policies and programs directed towards population management. This is in the premise that the population dynamics can either hinder or stimulate economic development. The theory emphasizes the importance of the demography of the population such as the age structure of the population, the labor force, the employment rate, and family well-being (Bloom et al., 2003). Population growth can directly and indirectly affect the economy, but economy can also directly and indirectly affect the population (Tsen and Furuoka, 2005) – forming a complex feedback system.

Policy-makers can use quantitative analysis in formulating a national roadmap for development. The Population White Paper by the Singaporean government (NPTD, 2013) is a good example of a roadmap. Population management involves realizing effective population-associated policies towards a balanced age structure, and implementing socio-economic development strategies to attain a sustainable level of population processes.

## OBJECTIVES OF THE STUDY

Using quantitative forecasting, this research aims to determine the future population structure of the Philippines if the government imposes a child policy. This research also aims to identify various implications of the child policy implementation.

In this paper, the authors define *n-child policy* (or simply, child policy) as a state rule imposing a maximum number of children per couple (in a strict sense), or a program that promotes an ideal maximum number of children per couple (in a lighter sense). The authors formulated numerical calculations to describe various "what-if" scenarios illustrating the consequences of the n-child policy (where n=0,1,2,3 or 4). The discussion focuses on quantitative insights arising from the various population growth scenarios. In addition, the authors sometimes refer to zero growth rate as synonymous to population stabilization; however, note that zero-growth is often transient.

## METHODOLOGY

The authors developed a program in Microsoft Excel platform to perform the computations involving age-structured compartmental discrete model for projecting population growth. Appendix Figure Set 1 presents diagrams illustrating the compartmental model. The program has multiple worksheets linked together and has three types of files: the input, the computation and the output files. This model is an alternative to the existing techniques, such as the cohort component method.

The following are the contents of the program. In this paper, the base year is 2011.

1) **The Input file**. This file contains the following data:

   a. *Current Population, with estimated age distribution*. The ages considered are from age 1 to 100.

   b. *Birth Rate*. In this paper, the authors define crude birth rate as the annual average births per 1000 people before considering the child policy. The authors used a regression model calculated in CurveExpert with a coefficient of determination $R^2 \approx 0.98$ to forecast the birth rates (see Figure 1). The birth rate function is a decreasing function that indicates a declining rate over time. In this paper, the authors assume a conservative age range for females giving birth, which is age 18 to 45.

   $$rate_{birth}(t) = \frac{3.66}{1 - 10e^{-1.07t}} \qquad (1)$$

   c. *Death Rate*. Crude death rate is the annual average deaths per 1000 people. This study employed a regression model calculated in CurveExpert with a coefficient of determination $R^2 \approx 1$ to forecast death rates. The death rate function is a decreasing function that indicates a declining death rate over time.

   $$rate_{death}(t) = \frac{4.21}{1 - 6.29e^{-1.76t}} \qquad (2)$$

   d. *Child Policy*. The user of the program can input the maximum number of children that a couple can have in a lifetime. The choices are 0 child, 1 child, 2 children, 3 children and 4 children. The user can also opt not to impose a child policy (no child policy or the status quo). The model incorporates the child policy on top of the crude birth rate. Hence, the effective birth rate is equal to the crude birth rate plus the effect of the child policy.

   e. *Sex Distribution*. The model employs a constant percent of male and female in a population.

   f. *Current Gross Domestic Product (GDP)*.

   g. *Coefficient of Variation for Sensitivity Analysis*. In this paper, the standard deviation in the sensitivity analysis is equal to the 10% of the original parameter value.

h. *Net Migration*. The authors assume a constant net migration since there are no suitable available historical data for curve fitting. The estimated population of the Philippines does not include Filipinos living in other countries. Migration greatly affects population reduction in the Philippines (Gaston, 2007).

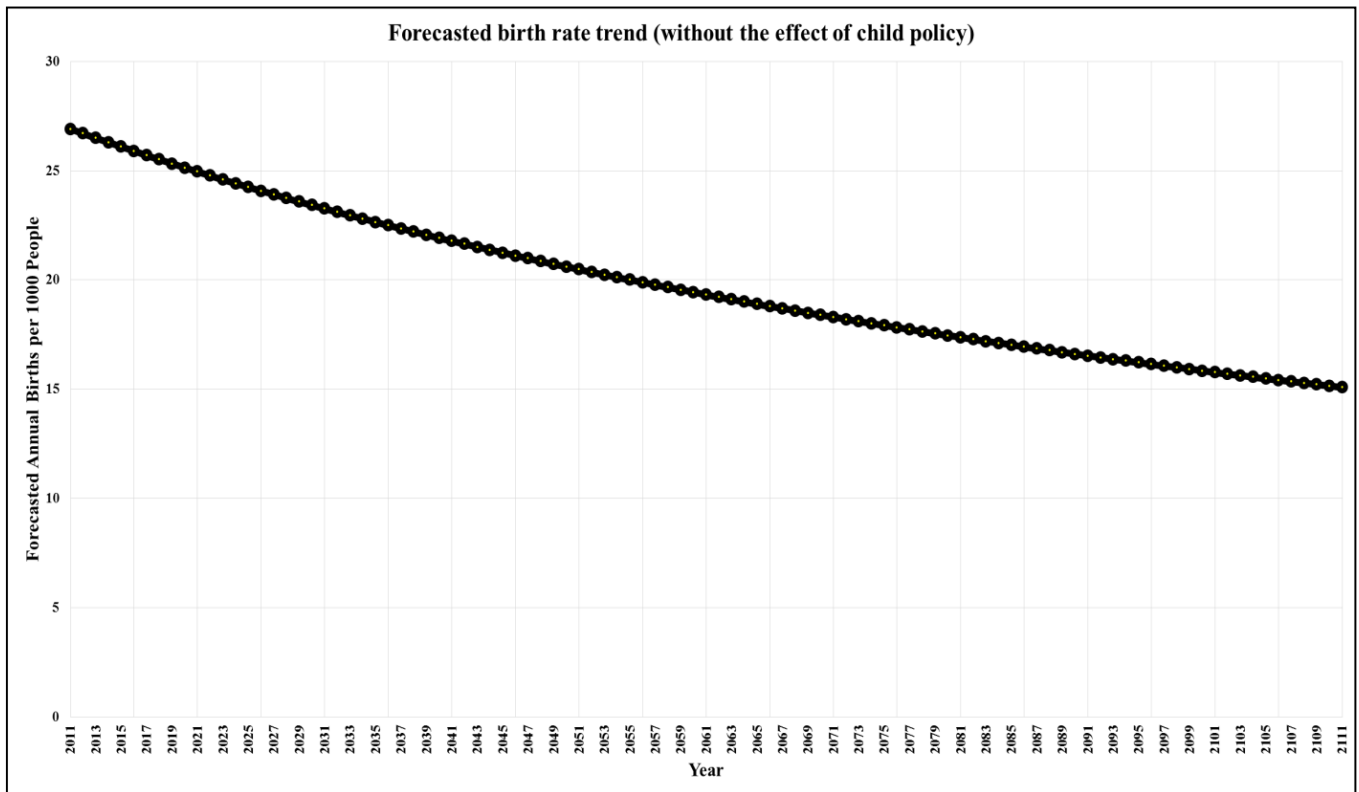

Figure 1. Projected average births per 1000 people

2) **The Output File (Graphs)**. The Results and Discussion part of this paper presents the generated graphs. The forecast time frame is 100 years.

3) **The Computation File**. The General Computation Worksheet performs the necessary computations in the program. The formulas used are available in the supplementary materials.

4) **Sensitivity Worksheets Files**. The worksheets are the same as the General Computation Worksheet, but with random numbers (i.e., birth, death and net migration rates are noisy parameters). Perturbation is per parameter value.

The Microsoft Excel program is a template for studying population dynamics. Future research can use the calculations as guide. Users can edit the worksheets to suit other objectives, and they can apply the template to other countries, as well. The Microsoft Excel files are accessible in the supplementary materials.

Figure 2. The Computation File

**RESULTS AND DISCUSSION**

This study shows several insights, possible challenges and consequences of the dynamics of Philippine population. Most of the data used in this paper are either historical or estimated data from various agencies (e.g., the National Statistics Office (NSO, 2012b), National Statistic Coordination Board (NSCB, 2012, 2013), World Bank (World Bank, 2013), Central Intelligence Agency (CIA, 2013), International Monetary Fund (IMF, 31) and Index Mundi (Index Mundi, 2013)). The discussion focuses on the behavior of the dynamics rather than the predicted values because the values may not necessarily be exact due to statistical errors in the raw data estimates.

Figures 3 and 4 show the effect of imposing a child policy with the influence of the declining birth and death rate regression curves. Controlling overpopulation does not have immediate effects in the macro-level. As shown in Figure 3, the effect of child policy starts to become visible on a macro-scale only after several years. Hence, long-term implementation is necessary before results become evident. The government should devise consistent strategies for attaining a sustainable population growth.

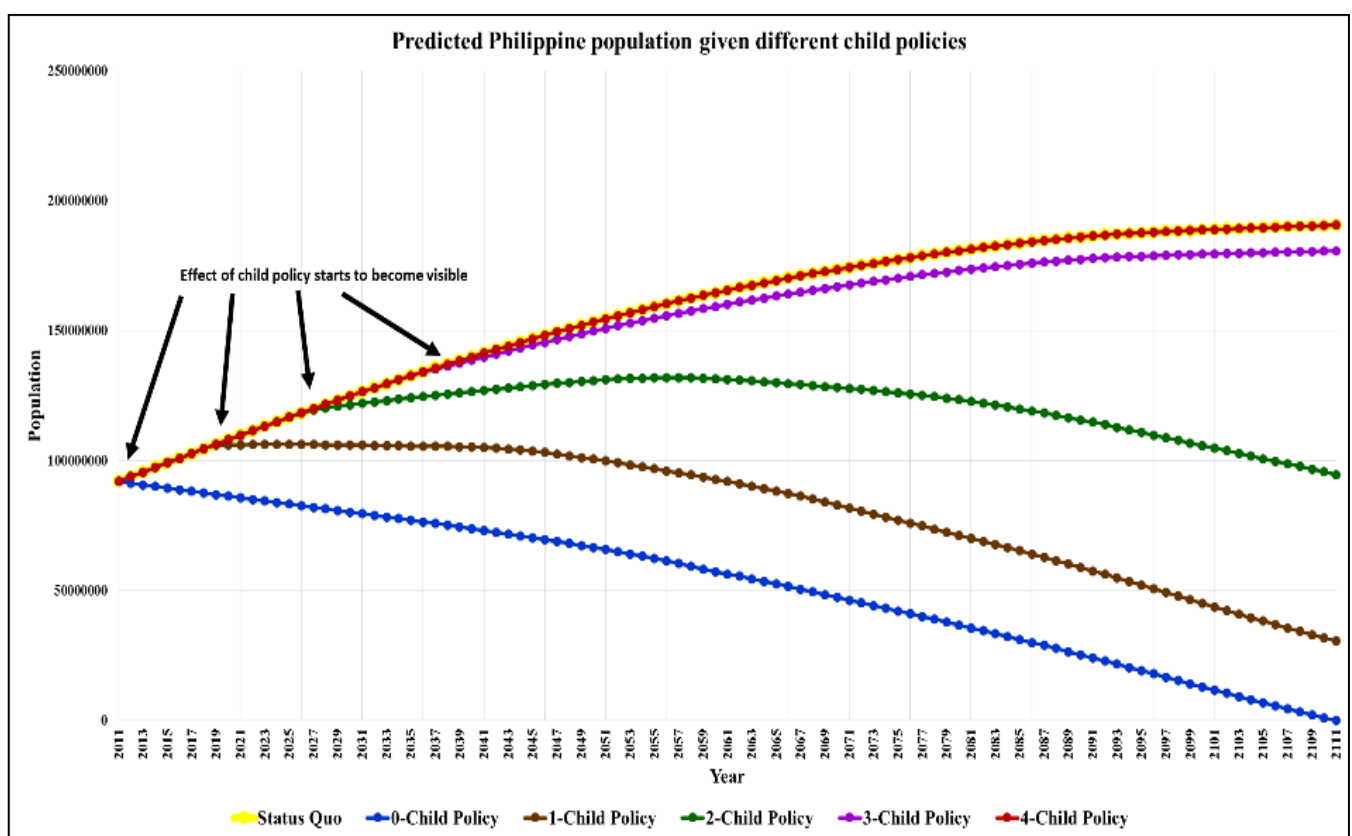

Figure 3. Projected population of the Philippines

Notice that the time series of the situation with four-child policy coincides with the status quo. This is due to the declining trend of the average birth rate of Filipinos which limits the number of newly born children per year. The decreasing trend of annual birth rate can be due to various social factors (e.g., socio-economic development in the Philippines). In the status quo or situation with a child policy dictating four or more children (as well as in the three-child policy), the population approaches zero growth rate. In the status quo, the population will not grow exponentially. This implies that regulating the average birth rate (or fertility rate) of Filipinos is an auxiliary way for population control. However, attaining zero growth rate through controlling birth rates without child policy is a long-term endeavor. Furthermore, note that when birth rate remains constant, the Philippine population does not converge to zero-growth in 100 years (see Figure 9).

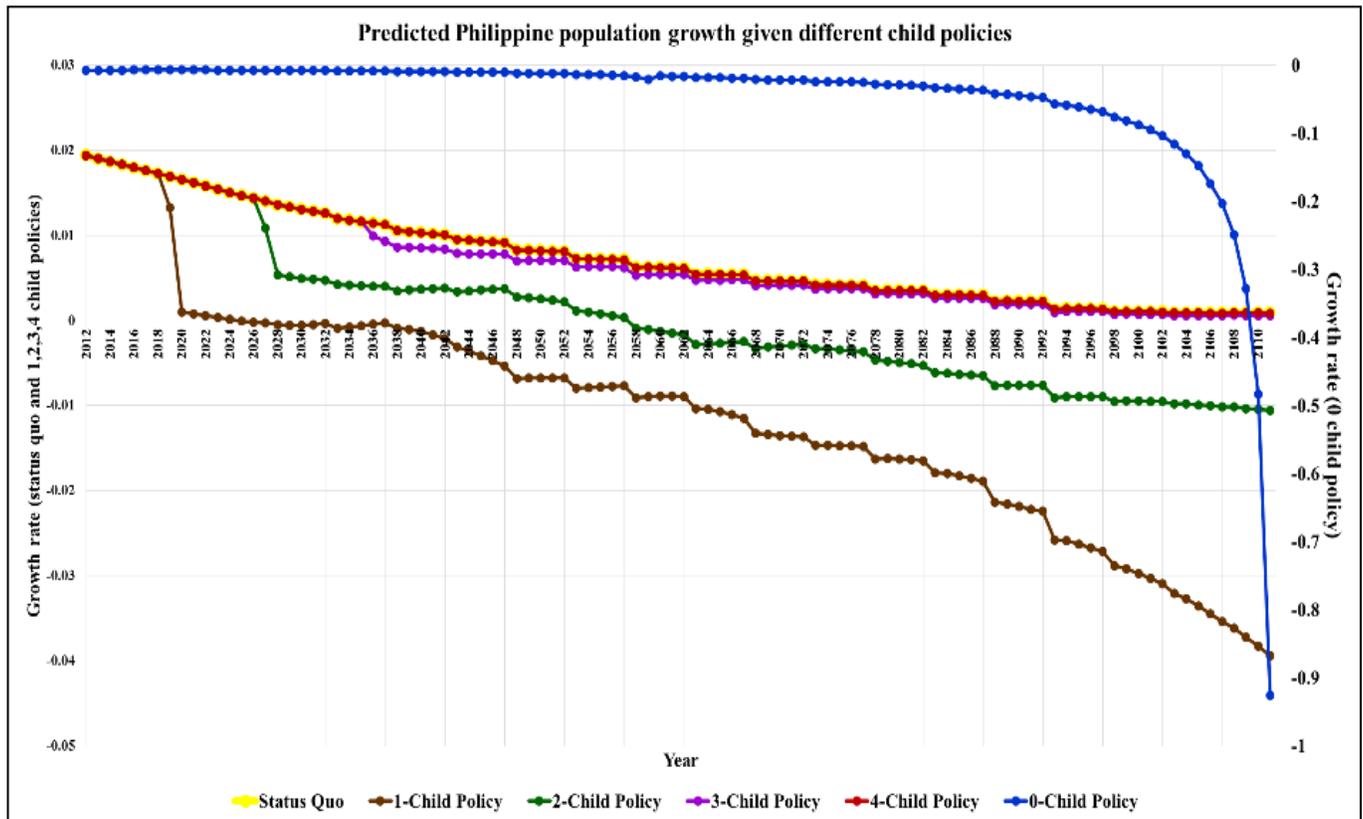

Figure 4. Projected growth rate of the Philippine population

Figure 5 illustrates the resulting age structures. Age structure of the population is very important to the economy of the country (Bloom et al., 2003). With one-child policy, the population will have an inverted age pyramid (ageing-dilemma). With two-child policy, the population tends to have a box-like age structure. In the status quo and situations with child policy dictating three or more children, the population will have a moderate pyramid-shaped age structure.

An ageing population (inverted pyramid) and juvenile-dominated population (perfect pyramid) put pressures on the labor force, such as in sustaining the social services and child care. The state should strive in pushing strategies that result to a longstanding beneficial age structure. The labor age group should be able to sustain the support both for the young and retired age groups.

Various existing policies and programs promote a maximum of two children per couple to have a balanced age structure. However, the age structure resulting from two-child policy is between the possibility of the inverted pyramid and regular pyramid scenarios. There is a danger that the national average fertility rate can go above or below the replacement level. Hence, the state should perpetually monitor the outcome of the two-child policy and provide insurance to minimize the associated risks.

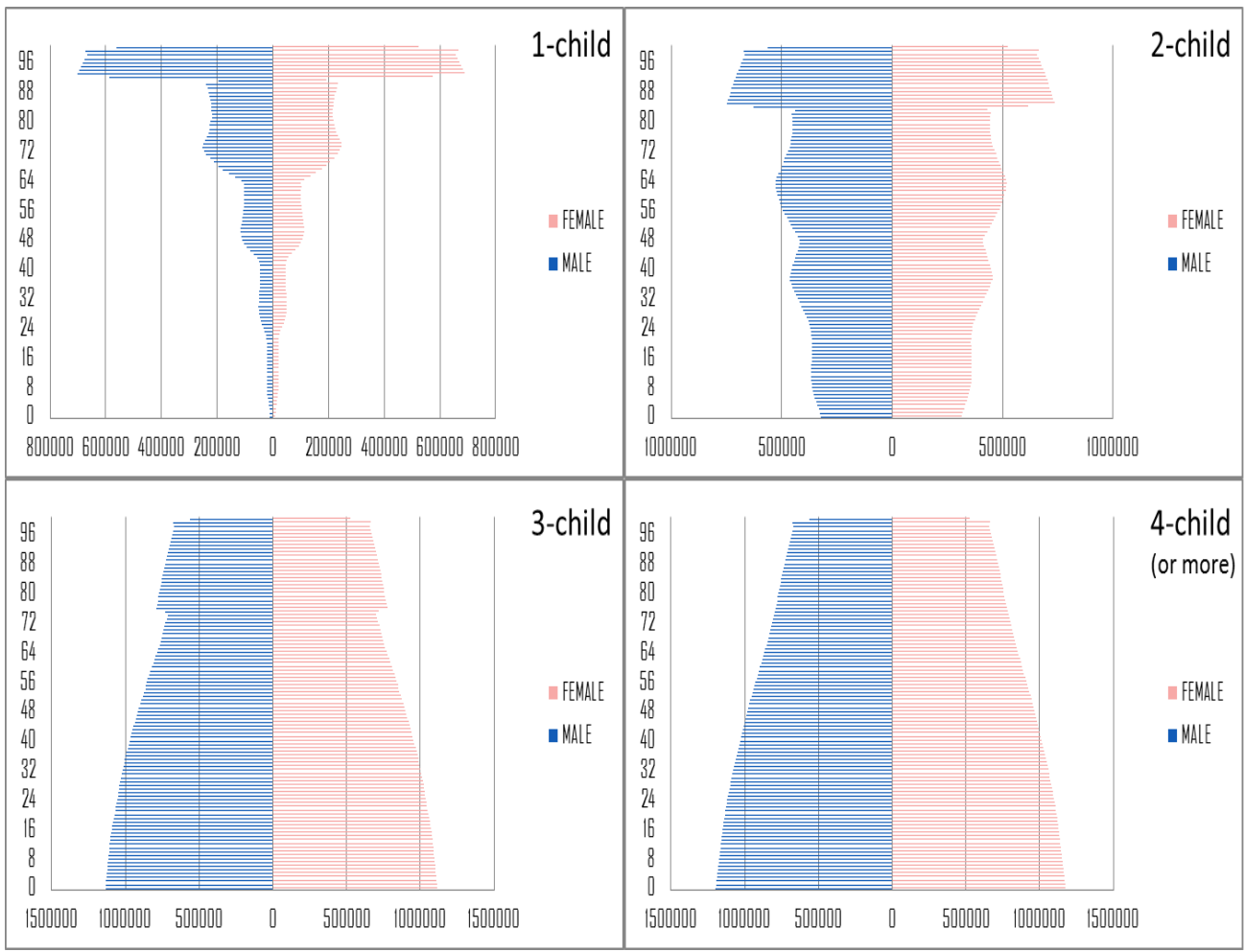

Figure 5. The projected age structure after 100 years

A consistent and balanced age structure is necessary for a productive and stable socio-economic growth. Figure 6 illustrates the estimated age group distribution in year 2036 per scenario. The figure shows the percentage of the young and prime labor force (age 18-59) that could support the juvenile (age 0-17) and ageing (age 60 plus) population. A large age group in a period can cause unbalanced population structure in the future. For example, the outcome of one-child policy shows that there would be a large number of people with ages 18-59 in 2036. This may result to a sizeable labor force in year 2036, but may lead to an ageing dilemma in the future. After 2036, there will be a massive number of people needing social services yet a relatively small number of people who would support the whole population.

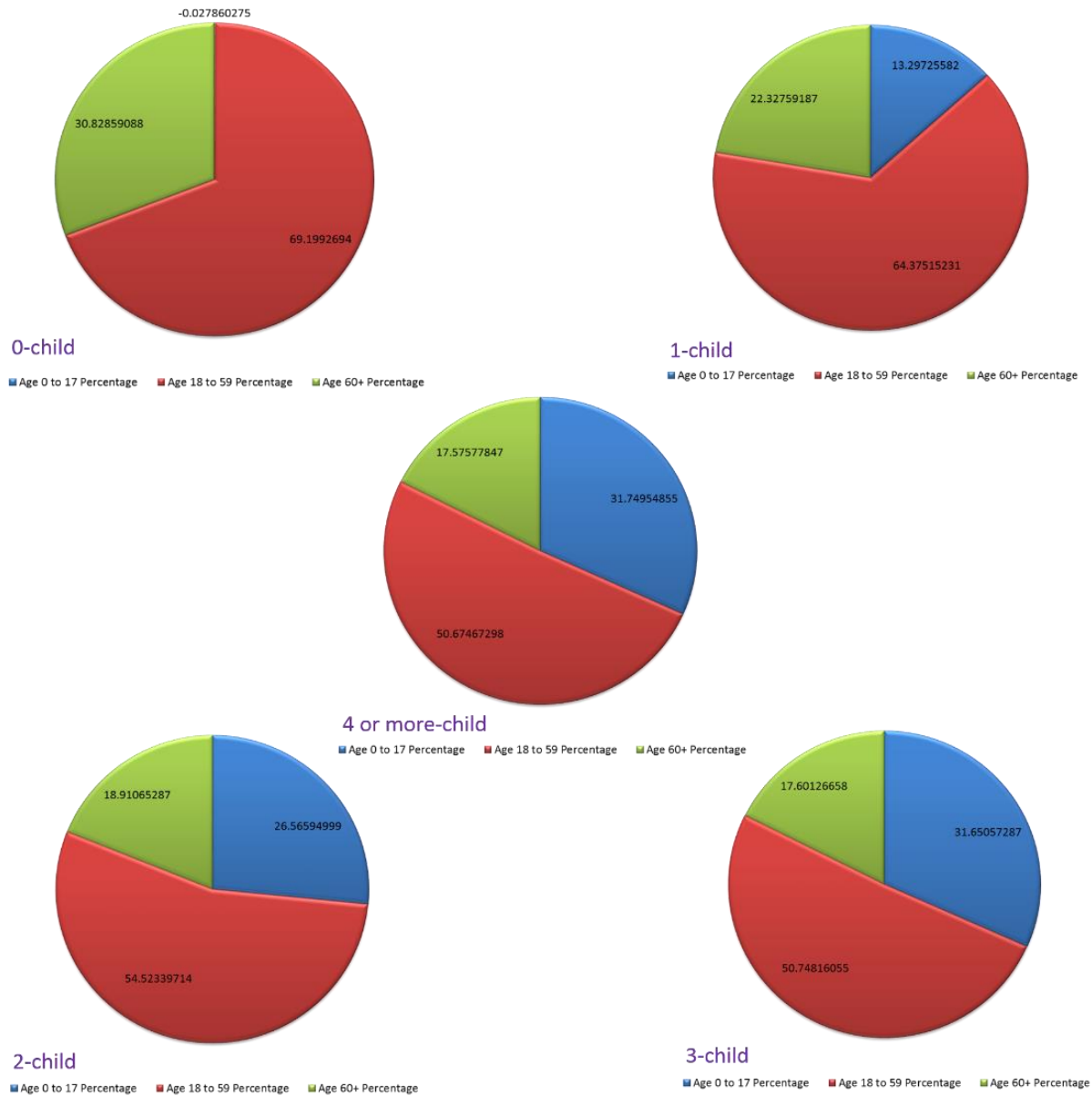

Figure 6. Age group distribution, 25 years after implementing the child policy

The population dynamics of the country may have an effect on the economy. Although, the effect can be two-way, that is, population growth affects the economy and the economy affects the population. In this paper, the focus is on the possible values of the per capita GDP growth given the different estimated population trends. Future research can consider other economic indicators. Table 1 presents the required annual GDP growth rate of the Philippines so that in year 2036, the per capita income of Filipinos is PhP 29,830 which is a 5% annual increase from 2011. Note that this forecast does not include income disparity among social classes.

Table 1. Required annual GDP growth rate to have 5% annual increase in per capita income

|  | 0-child | 1-child | 2-child | 3-child | 4 (or more)-child |
|---|---|---|---|---|---|
| Geometric mean for 25 years (%) | 4.22 | 5.58 | 6.28 | 6.59 | 6.59 |

*Estimated monthly GDP per capita in the year 2036 = PhP 29,830.*

In 2011 and 2012, the recorded GDP growth rates of the Philippines are around 3.6% and 6.8%, respectively (NSCB, 2012). The rates shown in Table 1 are achievable yet are relatively higher than the developed economies. Furthermore, the resulting PhP 29,830 per capita income is still far below the developed countries. In 2012, Japan and the United States have monthly nominal per capita income of approximately PhP 175,000 and PhP 187,000, respectively (at 1 USD = PhP 45) (IMF, 2013). There are ways to deal with this problem, such as considering an immoderate sloping downward birth rate trend and pushing for a progressive economy (to increase the GDP). However, these general solutions entail complex and detailed strategies. In addition, the policy-makers should investigate the micro-socioeconomic effects of population dynamics in detail. The effects of population growth may not be apparent at the macro-level, but the families (at the micro-level) may have already been experiencing the outcomes.

Using parameter estimation, the authors predicted the approximate logistic (Verhulst) growth rate and carrying capacity of the Philippines ($dP/dt = rP(1 - P/K)$; $P$ is population size). Based on historical data and results of the discrete computation, the approximate Verhulstian growth rate of the Philippine population is $r \approx 3.4632\%$ and the Verhulstian carrying capacity is $K \approx 198,873,000$ people (using tolerance error=0.001). The authors hypothesize that when the Philippine growth rate and population size converge to the estimated Verhulstian parameters, the Philippines might suffer from the harmful consequences of overpopulation. A population size near the carrying capacity is already at the edge of saturation. The population of the Philippines should be considerably lower than the logistic carrying capacity to abate the detrimental outcomes. However, this estimation uses time series data. Future research can compute the actual carrying capacity by approximating the total available resources. Moreover, since resources are not constant, future studies may consider a dynamic carrying capacity.

Figure 7 displays that the trend of the Philippine population has already undergone a change in concavity (already passed the inflection point) – from an increasing to a decreasing growth rate. This phenomenon might be due to several complex factors, such as socio-economic development. Another possible conjecture is that the population might have already started experiencing the pressure set upon by the carrying capacity.

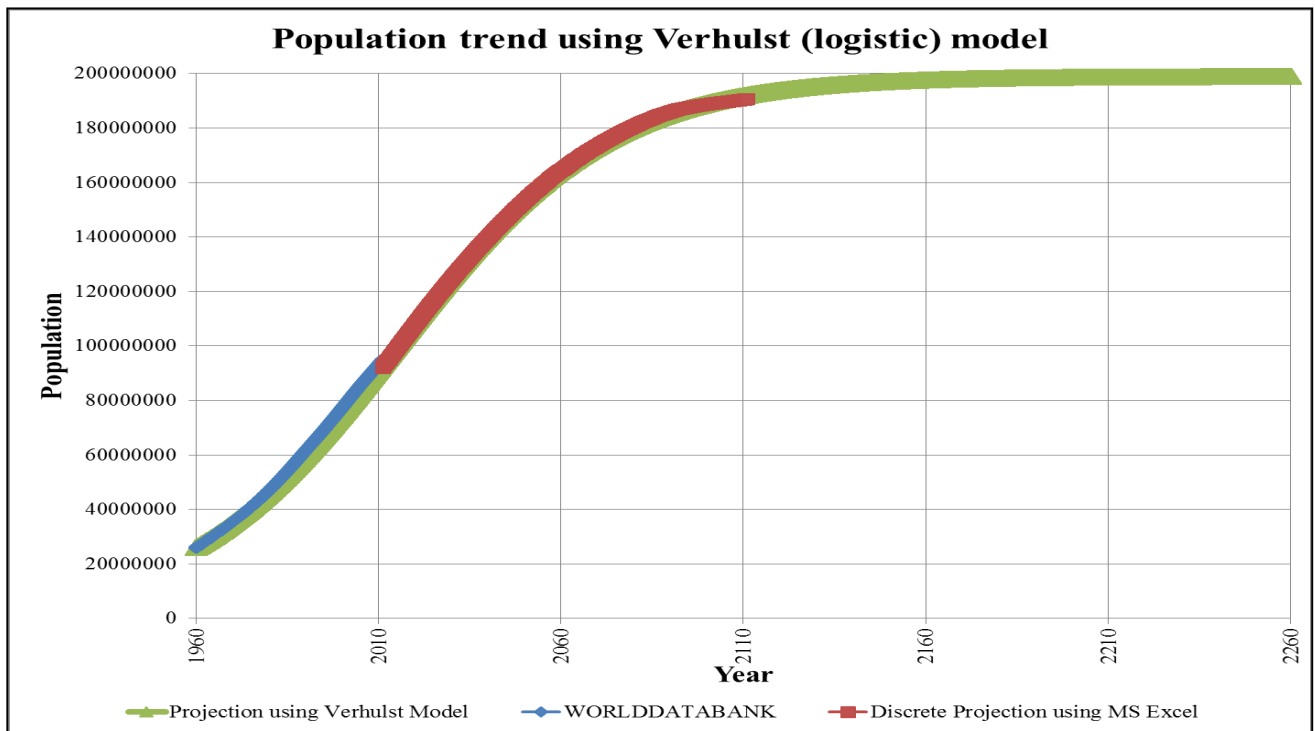

Figure 7. Population trend of the Philippines using Verhulst model. The population converges to the carrying capacity of 198,873,000 people

In addition, Figure 8 shows the trend of the population density from 2011 to 2111 (calculated using the total land area of the Philippines). The maximum population density for two-child policy is around 450 people per square kilometer; while, for the status quo, the population density can go beyond 600 people per square kilometer. This computation is at the macro-level perspective. For future research, population density outlook can be estimated per province, city or municipality.

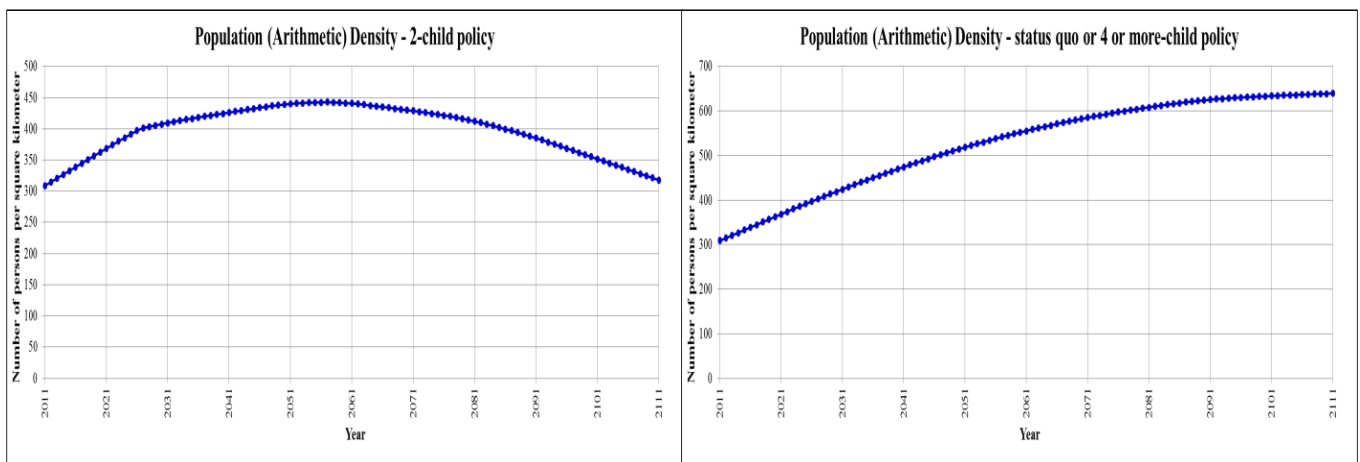

Figure 8. The projected population density (left is assuming 2-child policy; right is assuming status quo or 4 or more-child policy). The density is in terms of number of people per km$^2$

## The ideal child policy

One of the natural assets of the Philippines is its demography. The Philippine economy heavily relies on its population, such as the laborers that supply the service sectors. There is also a huge bulk of remittances from Overseas Filipino Workers. The setting of the Philippine economy and technology utilization may change in the future, but it is beneficial for the current policy-makers to formulate and implement long-term plans that address population processes (fertility, mortality and migration) leading to the desired socio-economic outcomes.

Figure 9 shows that when birth rate and death rate are constant, and there is no child policy, the Philippine population seemingly increases and does not stabilize in 100 years. This trend is detrimental to the Philippines when the country's limited resources do not match the ballooning population size.

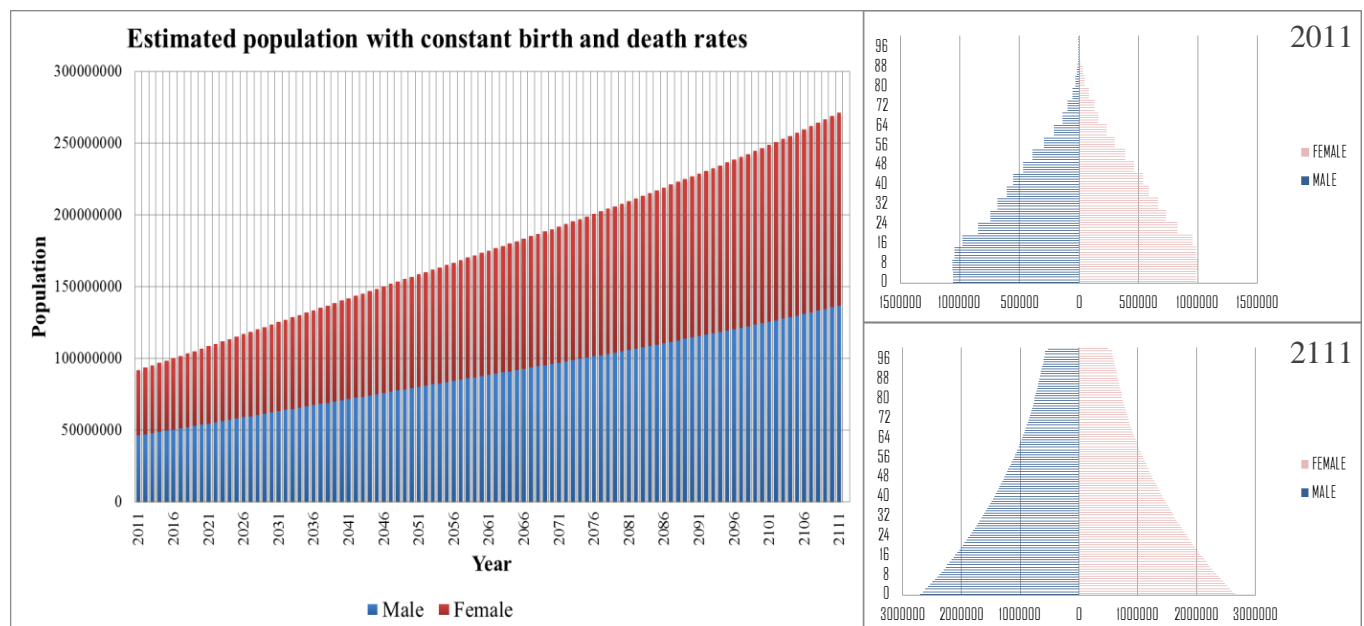

Figure 9. Projected population when crude birth rate = 24.807 (per 1000 people) and crude death rate = 5.753 (per 1000 people) are held constant and no child policy is implemented. Similar behavior can be seen if birth rate only is held constant

The question now is "what is the ideal maximum number of children per couple that the Philippine government should promote?" The answer may rely on the plans of the government about the desired future population structure of the country. This desired future population structure can be a national goal to achieve sustainable socio-economic development.

From the preceding discussions, it is apparent that the three and four-child policies coupled with the declining birth rate trend lead to potential population stabilization. However, if the population is unable to sustain the declining birth rate trend, the population dynamics scenario may change. Suppose the birth rate trend is constant at 24.807 per 1000 people. Then, the two-child and the three-child policies result to the population scenarios shown in Figure 10 and Figure 11, respectively.

In the model with constant birth rate, four-child policy does not result to a stabilizing population during the 2011 to 2111 period. It seems that the potential population stabilization materializes in a three-child policy (both in the models with declining birth rate trend and constant birth rate). Nevertheless, it is important to note that the stabilization happens upon reaching the population size of around 200M people, which is beyond the estimated Verhulstian carrying capacity. Moreover, stabilization is feasible only after one hundred years from the start of child policy implementation.

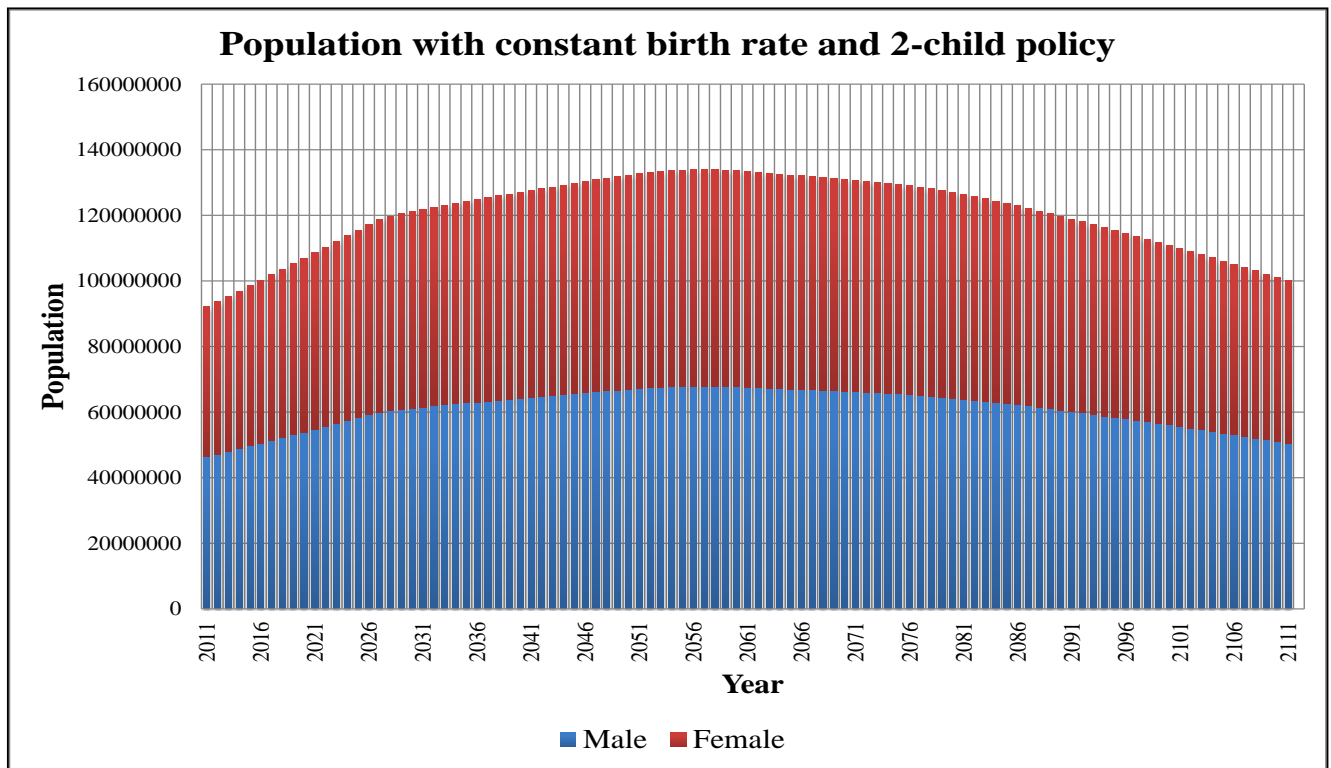

Figure 10. Projected population when crude birth rate = 24.807 (per 1000 people) is held constant and 2-child policy is implemented

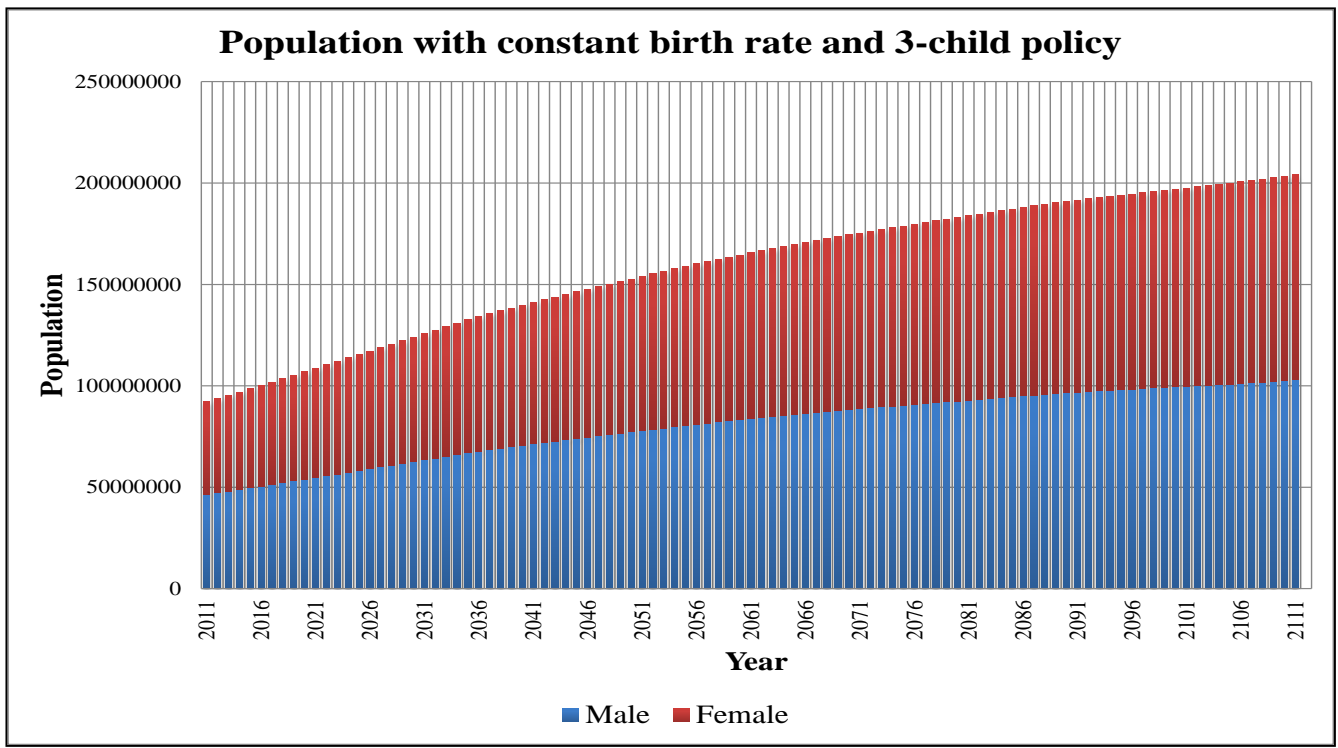

Figure 11. Projected population when crude birth rate = 24.807 (per 1000 people) is held constant and 3-child policy is implemented

Promoting a two-child policy can be favorable at the micro-socioeconomic and macro-socioeconomic levels, given that there is enough government support. Having two children per couple is ideal for forming a sustainable family size, especially when considering the finances for child care, education and daily sustenance (as compared to more than two children per couple). Long-term population stabilization may not be achievable in a two-child policy, but the population size and age structure (boxed-shape) can be sufficient to support the juvenile and senior population group for 100 years. Zero-growth rate is achievable in around 50 years of policy implementation, but afterwards the population will decline. In years 2011 to 2111, the population size will not increase beyond 140M people, but it will not decrease below the current 2011 population of around 90M. The expected population decline is not as drastic as the situation with one-child policy. Note that the historical trend and the projected trend with two-child policy collectively resembles demographic archetypes with sigmoidal curve (S-shaped) in the early stages but will soon have a declining population phase (Henslin, 2012).

The policy-makers may opt to promote a three-child policy depending on their plans and programs in establishing a sustainable socio-economic development. A three-child policy may lead to population stabilization, but the population may converge beyond the estimated Verhulstian carrying capacity. This is observable both in the models with declining birth rate trend and constant birth rate. The risk of converging beyond the carrying capacity can be resolved by increasing the value of the carrying capacity (available resources and infrastructure) of the Philippines. Proportionate socio-economic development (education, livelihood, housing, social services, health care, etc.) and environmental sustainability should back up this population growth. It is important to take into account the time duration before stabilization because zero growth rate may happen only after a century. Consequently, government strategies can harmonize the two-child and three-child policies depending on the results of the periodical reviews of the actual situation and policy outcomes.

**Reliability Analysis**

To check the reliability of the calculations, the authors performed sensitivity analysis using random perturbation of birth, death and migration rates. The authors also compared the results with the Philippine National Statistics Office (NSO) projections (NSO, 2013). The estimates are relatively acceptable and yield good prediction about the behavior of the population.

Figure 12 shows the increase in propagated relative error (the authors define relative error as the absolute difference between the original and the perturbed computation divided by the original computation) through time. This is discernible because the collective errors accumulate per time step. However, in most of the computations, the error only increases in a nearly linear manner (i.e., not a drastic increase), which is a reasonable intensity of error propagation. The model is robust against some degree of random noise.

Figure 13 presents the NSO projection (using medium assumption) against the forecast (using status quo computation or no policy implementation). The forecast closely coincides with the NSO estimates with Mean Absolute Percent Deviation = 2.73%. Moreover, the forecasts are within the range of NSO projection. For example, in 2040, the authors' estimate is 140,013,073 and the NSO projection is 132,536,300 (using low assumption) to 147,131,700 (using high assumption).

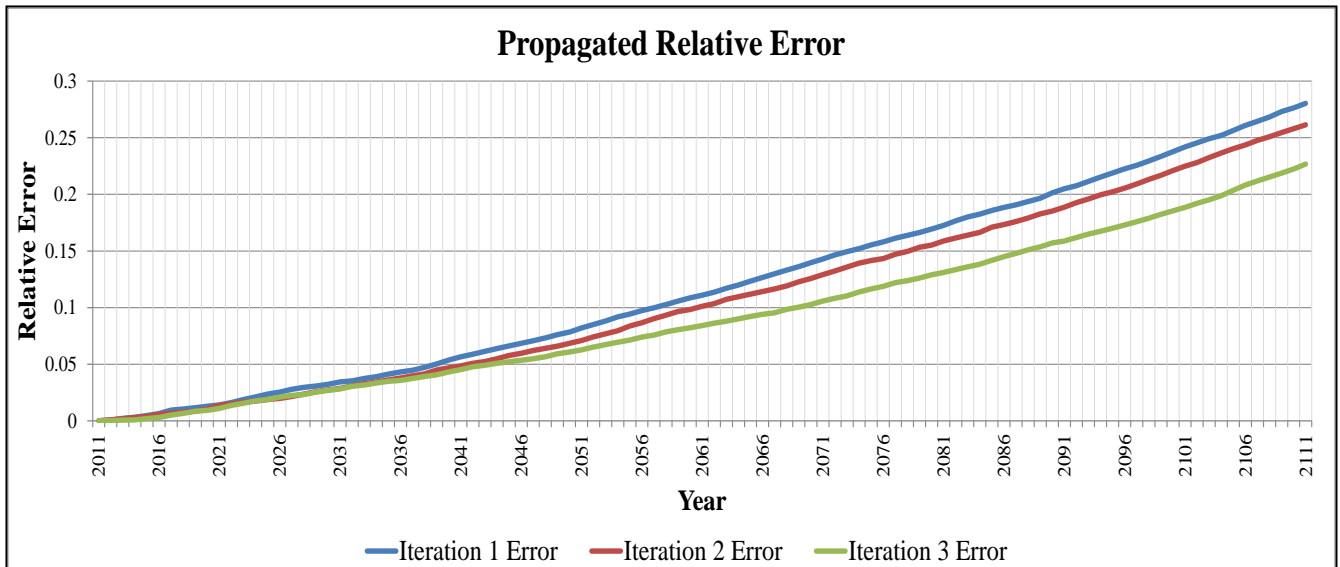

Figure 12. Propagated relative error where each parameter value is randomly perturbed with coefficient of variation=10%.

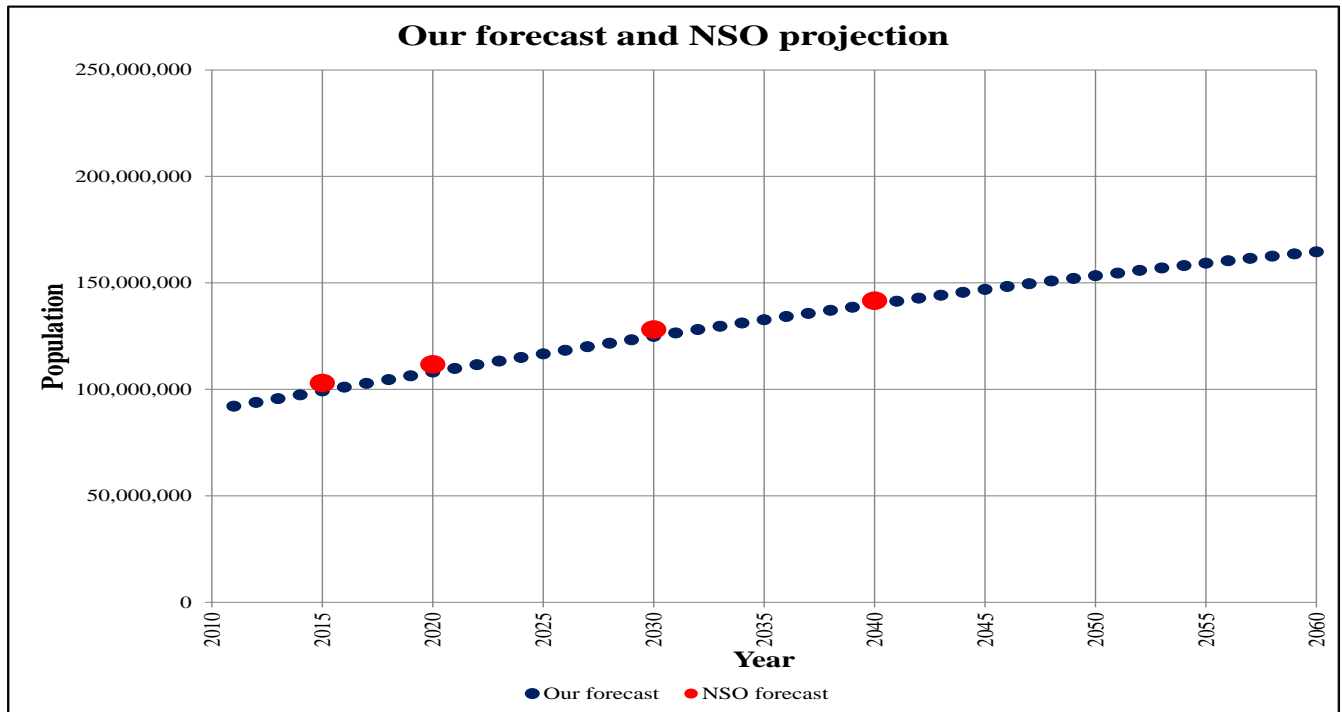

Figure 13. Comparison of the NSO population projection (using medium assumption) and the authors' forecast (using status quo computation).

## CONCLUSIONS

The forecast uses an age-compartment model with input data gathered from literature and databases. The aim is to investigate the different population structure scenarios resulting from the implementation of various child policies. Policies for population management can promote a maximum number of children per couple depending on the long-term goal of the national government. The creation, approval and implementation of policies can greatly influence the public's attitude toward national development agenda and social issues.

From the simulations, promoting either a two-child or a three-child policy is beneficial. However, these two policies produce different outcomes, and each has its own pros and cons. Within the forecast period from 2011 to 2111, a policy promoting a maximum of two children per couple may lead to a population decline after attaining zero-growth rate. A three-child policy may lead to stabilization yet may converge beyond the calculated Verhulstian carrying capacity, but saturating the carrying capacity can be resolved by increasing the available resources of the Philippines. A child policy dictating a maximum of four or more children per couple results to a similar population growth as the status quo as an effect of the current declining birth rate trend. Population stabilization can be attainable even without implementing a child policy because of the declining birth rate trend, but this is possibly realizable only after 100 years.

The projected outcomes will aid policy-makers in the formulation and implementation of development strategies grounded on historical data and mathematical prediction. The effect of imposing a child policy is likely to be observable on a macro-scale only after several years from the start of policy execution. Thus, long-term planning and effective implementation are necessary. Furthermore, continuous review of policy consequences is essential. As a caution, a zero-growth may not be a mathematically stable steady state (such as a declining population size may follow after attaining zero growth rate). It is also imperative to have a continuous monitoring of the age-structure of the Philippines.

Socio-cultural concerns, urbanization, sustainable natural resource management and population growth create complex interrelationships in a dynamic system. Considering complex factors, together with other unpredictable spatiotemporal consequences brought about by war and natural calamities, makes policy-making a very difficult task to perform. Nonetheless, population growth model based on historical data will, more or less, give a glimpse of the scenario that the Philippines might face regardless of these complex factors. Population models provide policy-makers with a workable framework for formulating policy roadmaps. Integrating more dynamic and complex variables (socio-economic, political and ecological) and local factors (e.g., by region, province or town) can be the subject matter of future interdisciplinary population studies.

## ACKNOWLEDGEMENT

The authors gratefully acknowledge Mark Jayson V. Cortez and the journal referees for reviewing the manuscript.

**NOTES**
DASJ Talabis and EJV Manay contributed equally to this work. JF Rabajante is the corresponding author. The appendices and supplementary materials are available at http://jfrabajante.weebly.com/supplementary-materials-1.html.

**Software used**